\documentclass[aps, 10pt, prd, twocolumn, superscriptaddress, preprintnumbers, longbibliography, nofootinbib]{revtex4-2}

\usepackage[top=2.8cm, bottom=2.8cm, left=2cm, right=2cm]{geometry}
\usepackage[utf8]{inputenc} 
\usepackage[T1]{fontenc}
\usepackage{amsmath, amssymb, lmodern} 
\usepackage{graphicx}
\usepackage{subfigure}
\usepackage{natbib}
\usepackage{booktabs}
\usepackage{hyperref}
\usepackage{tabularray}
\usepackage{orcidlink}
\hypersetup{colorlinks, citecolor = blue, filecolor = blue, linkcolor = blue, urlcolor = blue}
\usepackage[normalem]{ulem}

\newcommand{\dd}{\text{d}}
\newcommand{\ra}{R_\text{A}}
\newcommand{\dra}{\dot{R}_\text{A}}
\newcommand{\DE}{\text{DE}}
\newcommand{\HDE}{\text{HDE}}

\begin{document}

\title{Inconsistencies of Tsallis Cosmology within \\ Horizon Thermodynamics and Holographic Scenarios}

\author{Pedro M. Ibarbo-Perlaza.\orcidlink{0000-0001-7374-7613}}
\email{Contact author: pedro.ibarbo@correounivalle.edu.co}
\affiliation{Departamento de Física, Universidad del Valle, 760032 Cali, Colombia}

\author{J. Bayron Orjuela-Quintana\orcidlink{0000-0001-5090-2860}}
\affiliation{Departamento de Física, Universidad del Valle, 760032 Cali, Colombia}

\author{Jose L. Palacios-C\'ordoba}
\affiliation{Departamento de Física, Universidad del Valle, 760032 Cali, Colombia}

\author{C\'esar A. Valenzuela-Toledo\orcidlink{0000-0002-6380-2778}}
\affiliation{Departamento de Física, Universidad del Valle, 760032 Cali, Colombia}

\begin{abstract}
We investigate the cosmological implications of Tsallis entropy in two widely discussed frameworks: the Cai–Kim thermodynamic derivation of the Friedman equations and the Tsallis holographic dark energy (HDE) scenario, considering both the Hubble scale and the Granda–Oliveros (GO) cutoff as infrared regulators. In both cases, the dynamics introduce a nonextensivity parameter $\delta$, with the standard Bekenstein–Hawking entropy–area relation recovered for $\delta = 1$. While previous studies have suggested that only small deviations from extensivity are observationally allowed, typically requiring $|1 - \delta| \lesssim 10^{-3}$, here we go further and perform a systematic consistency analysis across the entire expansion history.

We show that even mild departures from $\delta = 1$ lead to pathological behavior in the effective dark energy sector: its density can become negative or complex, its equation of state may diverge, or it can contribute an unacceptably large early-time fraction that spoils radiation domination and violates BBN and CMB constraints. Our results sharpen and unify earlier hints of tension, providing a clear physical interpretation in terms of corrections that grow uncontrollably with the expansion rate toward the past.

We conclude that within both the Cai–Kim and HDE formulations, a viable cosmology emerges only in the extensive limit, effectively reducing the models to $\Lambda$CDM. More broadly, our findings emphasize the importance of dynamical consistency and cosmological viability tests, when assessing nonextensive entropy formalisms as potential frameworks for describing the Universe’s dynamics.
\end{abstract}

\maketitle

\section{Introduction}
\label{Sec: Intro}
The discovery of cosmic acceleration~\cite{SupernovaSearchTeam:1998fmf, SupernovaCosmologyProject:1998vns} has motivated extensive efforts to extend the standard cosmological model in order to explain the observed late-time expansion without invoking a cosmological constant of unknown origin~\cite{Weinberg:1988cp, Martin:2012bt, CANTATA:2021asi}. Among the different approaches, thermodynamic interpretations of gravity provide an appealing framework~\cite{Jacobson:1995ab, Eling:2006aw, Padmanabhan:2009vy, Verlinde:2010hp, Padmanabhan:2012ik}, wherein the Friedman equations can emerge from the Clausius relation applied to the apparent horizon, with the underlying entropy–area law determining the dynamics of the Universe~\cite{Cai:2005ra, Akbar:2006kj, Akbar:2006mq}.

While the Bekenstein–Hawking entropy, proportional to the horizon area~\cite{Bekenstein:1973ur, Hawking:1975vcx}, successfully reproduces the $\Lambda$ cold dark matter ($\Lambda$CDM) paradigm, several nonextensive generalizations have been proposed to account for possible deviations at large scales. A prominent example is the Tsallis entropy~\cite{Tsallis:1987eu}, which introduces a nonextensivity parameter $\delta$ quantifying departures from additivity, with the standard extensive case recovered for $\delta=1$. This proposal has been explored in a wide range of contexts, including inflationary dynamics~\cite{Teimoori:2023hpv, Keskin:2023ngx, Khodam-Mohammadi:2024iuo, Odintsov:2022qnn}, gravitational waves~\cite{Jizba:2024klq}, black hole thermodynamics~\cite{Tsallis:2013bhs}, the large-scale structure formation process~\cite{daSilva:2020bdc, Sheykhi:2022gzb}, late-time cosmic acceleration~\cite{Sheykhi:2018dpn, Barboza:2014yfe, Mendoza_Mart_nez_2024}, with the additional motivation that it might help to alleviate cosmological tensions such as those in $H_0$ and $\sigma_8$~\cite{Basilakos:2023kvk}.

Cosmological models based on Tsallis horizon entropy have been developed within different thermodynamic frameworks. In this work, we focus on two of them: the Cai–Kim approach to thermodynamic gravity~\cite{Cai:2005ra} and the holographic dark energy (HDE) scenario~\cite{Wang:2016och}, based on the so-called holographic principle~\cite{tHooft:1993dmi, Susskind:1994vu}.

The Cai--Kim approach reformulates the first law of thermodynamics on the apparent horizon, linking the energy flux crossing the horizon to variations in its entropy~\cite{Cai:2005ra}. When the Tsallis entropy is implemented in this framework, the resulting modified Friedman equations for the expansion rate $H$ acquire additional terms proportional to $H^{2(2-\delta)}$, thereby modifying the cosmic expansion history. This formulation has been considered as a possible mechanism to drive late-time acceleration~\cite{Lymperis:2018iuz, Asghari:2021lzu}, yet its impact on the early Universe remains largely unexplored.

A related framework emerges within the holographic dark energy (HDE) paradigm, in which the dark energy density is determined by an infrared (IR) cutoff and the entropy–area relation. When the Tsallis entropy is incorporated into this setup, the resulting holographic dark energy density acquires a characteristic nonextensive dependence on the Hubble parameter, analogous to that found in the Cai–Kim approach~\cite{Tavayef:2018xwx, Saridakis:2018unr}.

In the holographic context, the IR cutoff has a cosmological origin, and several prescriptions have been proposed in the literature~\cite{Nojiri:2005pu, Nojiri:2017opc, Gao:2007ep, Nojiri:2021iko, Nojiri:2021jxf, Nojiri:2022dkr, Nojiri:2022aof}. The most common choice identifies the IR cutoff with the Hubble horizon. However, this formulation suffers from the so-called causality problem: the present value of the dark energy density appears to depend on the future evolution of the scale factor, thereby violating causality~\cite{Li:2004rb}. This issue can be circumvented by adopting the Granda–Oliveros (GO) cutoff, originally proposed in Refs.~\cite{Granda:2008dk, Granda:2008tm} based on dimensional arguments. The GO cutoff successfully avoids the causality problem while maintaining the holographic consistency of the model. In what follows, we employ both cutoffs to investigate the Tsallis holographic dark energy scenario.

The central issue, as we demonstrate in this paper, is that the departures brought about by these thermodynamic-inspired alternatives cannot be treated as small perturbations around $\Lambda$CDM. Although the modified equations can be formally expressed as the $\Lambda$CDM background plus $\mathcal{O}(\delta-1)$ contributions, these terms scale as positive powers of $H$. As the Hubble parameter grows toward the past, the corrections inevitably dominate during radiation and matter domination, even for arbitrarily small deviations from extensivity. Consequently, the models exhibit unavoidable pathologies in the effective dark energy sector, such as negative density, divergent equation of state, and an excessive early contribution that disrupts the standard radiation--matter sequence. In this sense, the nonextensivity parameter $\delta$ does not provide a controlled deformation of $\Lambda$CDM, but instead destabilizes the cosmological background whenever $\delta \neq 1$.

Our work sharpens previous indications of tension in Tsallis-based cosmology by tracing these inconsistencies to their fundamental origin: the nonperturbative growth of the $H^{2(2-\delta)}$ corrections in the early Universe. The resulting constraints are extremely stringent, effectively casting doubts on viability of the Tsallis formulation—within both the Cai–Kim and HDE frameworks—as a compelling alternative to the standard model of cosmology.

This article is organized as follows: Sec.~\ref{Sec: Tsallis Cosmology} reviews the Tsallis entropy and its implementation within the Cai--Kim thermodynamic approach to gravity. In Sec.~\ref{Sec: Cosmo Evolution}, we analyze the cosmological evolution of this model and present our analytical derivation and numerical results, highlighting the inconsistencies that arise in the Cai--Kim formulation. In Sec.~\ref{Sec: Tsallis as a Perturbation}, we interpret these issues as uncontrolled perturbative corrections to the $\Lambda$CDM model. Sec.~\ref{Sec: THDE} discusses the Tsallis holographic dark energy scenario and shows that analogous pathologies also appear in this framework. Finally, Sec.~\ref{Sec: Conclusions} summarizes our findings and discusses their implications for nonextensive horizon thermodynamics.

\section{Tsallis Cosmology from Horizon Thermodynamics}
\label{Sec: Tsallis Cosmology}

In the thermodynamic–gravity approach as described by Cai--Kim in Ref.~\cite{Cai:2005ra}, the Friedman equations governing the evolution of the Universe are derived from the first law of thermodynamics (Clausius relation):
\begin{equation}
\label{Eq: Clasius relation}
    \delta Q = T_\text{h}~\dd S_\text{h},
\end{equation}
where $\delta Q$ denotes the heat flow across the cosmological horizon, $T_\text{h}$ is the associated temperature, and $\dd S_\text{h}$ is the variation of horizon entropy. This framework provides a natural arena to test non-standard entropies in a cosmological setting. In the following subsections, we will dissect this expression in order to unveil its cosmological consequences for the Tsallis entropy.

\subsection{Apparent Horizon}
In the cosmological context, the horizon is defined by the \emph{apparent horizon}~\cite{Frolov:2002va}, which is the marginally trapped surface with vanishing expansion. For a homogeneous and isotropic Universe described by the spatially flat Friedman-Lema\^itre-Robertson-Walker (FLRW) metric, the line element in spherical coordinates $(t, r, \theta, \phi)$ reads:
\begin{equation}
    \dd s^2 = - \dd t^2 + a^2(t) \left\{\dd r^2 + r^2\dd \theta^2 + r^2 \sin^2 \theta~\dd \phi^2 \right\}.
\end{equation}
Writing the metric as:
\begin{equation}
    \dd s^2 = h_{a b} \dd x^a \dd x^b + R^2 (\dd \theta^2 + \sin^2 \theta~\dd\phi^2),
\end{equation}
where $R = a(t) r$ is the physical radius, $a$ is the scale factor, and the two-dimensional metric in the $(t, r)$ plane takes the form $h_{a b} = \text{diag}(-1, a^2)$. From this metric, the apparent horizon radius $\ra$ satisfies:
\begin{equation}
    h^{a b} \left(\partial_a \ra~\partial_b \ra \right) = 0 \quad \Rightarrow \quad \ra = \frac{1}{H},
\end{equation}
where $H \equiv \dot{a}/a$ is the Hubble parameter. This is the familiar result that the apparent horizon corresponds to the Hubble horizon.

\subsection{Heat Flux and Horizon Temperature}

Following Ref.~\cite{Cai:2005ra}, the heat flow through the apparent horizon during an infinitesimal interval $\dd t$ is:
\begin{equation}
    \delta Q = A \, (\rho + p) \, H\ra \, \dd t,
\end{equation}
where $A = 4\pi \ra^2$ is the horizon area and $\rho$, $p$ denote the density and pressure of a perfect fluid.

The temperature associated with the apparent horizon is defined in terms of its surface gravity $\kappa$ as:
\begin{equation}
    T_\text{h} \equiv \frac{|\kappa|}{2 \pi}. 
\end{equation}
In stationary spacetimes, this reduces to the simple form $T_\text{h}=1/(2\pi \ra)$. In a cosmological setting, however, the apparent horizon is dynamical and its radius evolves with time. This naturally raises the question of whether the temperature definition should include additional corrections accounting for this time dependence.

In dynamical spacetimes, the surface gravity can be expressed through the Hayward–Kodama relation~\cite{Hayward:1998ee}:
\begin{equation}
    \kappa \equiv \frac{1}{2}\nabla^a\nabla_a \ra \quad \Rightarrow \quad 
    \kappa = - \frac{1}{\ra} \left( 1 - \frac{\dra}{2H \ra} \right).
\end{equation}
Although formally consistent, this expression complicates the thermodynamic derivation of the Friedman equations. A common resolution is to invoke an analogy with stationary black holes: small perturbations of the horizon induced by infinitesimal changes in mass do not require explicit corrections to the temperature in the first law of black hole thermodynamics~\cite{Wald:1993nt, Gao:2001ut}. By the same reasoning, in the cosmological context one often neglects the time variation of $\ra$ during an infinitesimal heat flow, treating the horizon radius as approximately fixed. This argument leads to the so-called quasi-static approximation:
\begin{equation}
\dra \ll 2 H \ra,
\end{equation} 
which is equivalent to requiring:
\begin{equation}
    \frac{\dra}{2 H \ra} \ll 1 \quad \Rightarrow \quad 
    \epsilon \equiv - \frac{\dot{H}}{H^2} \ll 2.
\end{equation}
The condition is indeed satisfied during a quasi–de Sitter phase with nearly constant $H$, but it clearly fails in the radiation-dominated ($\epsilon = 2$) and matter-dominated ($\epsilon = 3/2$) eras. Therefore, the quasi-static assumption appears conceptually misleading if applied universally.

Fortunately, such an approximation is not strictly necessary. As shown in Ref.~\cite{Cai:2008gw}, the apparent horizon of an FLRW universe possesses a well-defined Hawking temperature, fully analogous to the event horizon of a black hole, and given simply by:
\begin{equation}
    T_\text{h} = \frac{1}{2\pi \ra}.
\end{equation}

\subsection{Tsallis Entropy on the Horizon}
To close the system, we need to describe the entropy variation of the apparent horizon. From black hole thermodynamics, it is known that the horizon entropy is related to its area. The most famous case corresponds to the Bekenstein-Hawking relation in which $S \propto A$~\cite{Bekenstein:1973ur, Hawking:1975vcx}. However, it is known that for large-scale gravitational systems nonextensive (non-additive) entropies may be more appropriate. Within this category, the Tsallis entropy replaces the Bekenstein-Hawking relation by~\cite{Tsallis:1987eu}:
\begin{equation}
    S_\text{T} \equiv \frac{\tilde{\alpha}}{4 G} A^\delta,
\end{equation}
where $\delta$ quantifies the degree of nonextensivity ($\delta = 1$ recovers the Bekenstein-Hawking case) and $\tilde{\alpha}$ is a constant with appropriate dimensions.

\subsection{Modified Friedman Equation}
Combining the above ingredients in the Clausius relation~\eqref{Eq: Clasius relation}, we get:
\begin{equation}
    4 \pi G (\rho + p) H~\dd t = \tilde{\alpha}\delta~(4 \pi)^{\delta - 1}~\ra^{(2 \delta - 5)}\dra~\dd t.
\end{equation}
and using the continuity equation for a perfect fluid:
\begin{equation}
    \dot{\rho} + 3 H (\rho + p) = 0, 
\end{equation}
we can integrate to obtain the modified Friedman equations. The first equation takes the form:
\begin{equation}
\label{Eq: First Friedman Eq}
    H^2 = \frac{8 \pi G}{3} (\rho + \rho_\DE), 
\end{equation}
with an effective dark energy density:
\begin{equation}
\label{Eq: DE density}
    \rho_\DE = \frac{3}{8 \pi G} \left\{ H^2 \left[ 1 - \dfrac{\alpha\delta}{2 - \delta} H^{2(1-\delta)} \right] + \dfrac{\Lambda}{3} \right\},
\end{equation}
with $\Lambda$ as an integration constant. Differentiating and using the continuity equation yields the second Friedman equation:
\begin{equation}
    \dot{H} = -4 \pi G (\rho + p + \rho_\DE + p_\DE).
\end{equation}
where we have identified the dark energy pressure as:
\begin{align}
\label{Eq: DE pressure}
    p_\DE = - \frac{1}{8 \pi G} &\bigg\{  3 H^2 \left[ 1 - \frac{\alpha\delta}{2 - \delta} H^{2(1 - \delta)} \right] \nonumber \\
    &+2 \dot{H} \left[ 1 - \alpha \delta~H^{2(1 - \delta)} \right] + \Lambda \bigg\}.
\end{align}
Finally, the effective dark energy fluid can be completely characterized by its equation of state, $w_\DE \equiv p_\DE/\rho_\DE$:
\begin{equation}
\label{Eq: wDE}
    w_\DE = - 1 - \dfrac{2 \dot{H} \left[ 1 - \delta \alpha H^{2(1 - \delta)} \right]}{3 H^2 \left[ 1 - \dfrac{\delta \alpha}{2 - \delta} H^{2(1 - \delta)} \right] + \Lambda}.
\end{equation}
These expressions coincide with those obtained in Ref.~\cite{Lymperis:2018iuz}.

It is important to emphasize that the Tsallis modification introduces powers of $H^{2(1-\delta)}$ into the effective energy density and pressure. As will be shown in the next section, even tiny deviations from $\delta=1$ produce corrections that grow uncontrollably toward the past, eventually dominating over matter and radiation. This feature provides the fundamental mechanism behind the cosmological inconsistency of the model.

\section{Cosmological evolution}
\label{Sec: Cosmo Evolution}
Following Ref.~\cite{Lymperis:2018iuz}, the cosmological dynamics emerging from Tsallis horizon entropy can be described in a fully analytical manner. We begin by introducing the standard dimensionless density parameters:
\begin{equation}
    \Omega_i \equiv \frac{8 \pi G}{3 H^2} \rho_i, 
\end{equation}
with $i = m, r, \DE$, denotes matter, radiation, and the effective dark energy component, respectively. The first Friedman equation imposes the usual constraint:
\begin{equation}
    \Omega_r + \Omega_m + \Omega_\DE = 1,
\end{equation}
from which the Hubble parameter can be expressed as:
\begin{equation}
\label{Eq: H(z)}
    H(z) = H_0 \sqrt{\frac{\Omega_{m_0}(1+z)^3 + \Omega_{r_0}(1+z)^{4}}{1 - \Omega_\DE(z)}},
\end{equation}
where $\Omega_{i_0}$ are the present-day density parameters for matter and radiation, and $\Omega_\DE(z)$ gives the contribution of the nonextensive dark energy sector. Here we have adopted redshift $z$, as the time variable, with $1+z = 1/a$.

Given Eq.~\eqref{Eq: H(z)}, the other density parameters can be expressed in terms of $\Omega_\DE(z)$. In particular, the radiation density parameter reads:
\begin{equation}
    \Omega_r (z) = \frac{ \Omega_{r_0} (1+z) [1 - \Omega_\DE(z)] }{\Omega_{m_0} + \Omega_{r_0} (1+z)}. 
\end{equation}
Therefore, the system is closed if we are able to find an analytical expression for $\Omega_\DE$ in terms of $z$. In that way, computing the DE density parameter using the \DE~density in Eq.~\eqref{Eq: DE density} and the expression for the Hubble parameter in Eq.~\eqref{Eq: H(z)}, we obtain:
\begin{equation}
\label{Eq: DE Density Param}
    1 - \Omega_\DE= H_{(mr)}^2 \left\{ \frac{2 - \delta}{\alpha~\delta} \left[ \frac{\Lambda}{3} + H_{(mr)}^2 \right] \right\}^{\frac{1}{\delta - 2}},
\end{equation}
where we have defined:
\begin{equation}
    H_{(mr)}^2 \equiv H_{0}^2 \left\{\Omega_{m_0} (1 + z)^3 + \Omega_{r_0} (1 + z)^4 \right\}.
\end{equation}
The integration cosmological constant takes the form: 
\begin{equation}
\label{Eq: Cosmo constant}
    \Lambda = \frac{3 \alpha \, \delta}{2 - \delta} H_{0}^{2(2 - \delta)} - 3 H_{0}^2 (\Omega_{m_0} + \Omega_{r_0}).
\end{equation} 
This generalization captures the deviation from a constant vacuum energy due to the nonextensive nature of the horizon entropy.

\subsection{Conditions for Accelerating Expansion}
We next explore the parameter region in which the model produces late-time acceleration. The deceleration parameter is defined as:
\begin{equation}
    q \equiv - \frac{a\ddot{a}}{\dot{a}^2} = -1 - \frac{\dot{H}}{H^2}.
\end{equation}
and cosmic acceleration occurs for $q<0$. 

First, note that the parameter $\alpha$ has dimensions $[L^{2(1-\delta)}]$, where $L$ denotes a length scale. Therefore, we can absorb the parameter $H_0$ in terms of $\alpha$ and $\Lambda$ by redefining these constants as:
\begin{equation}
    \alpha \equiv \hat{\alpha} \, H_0^{2(\delta -1)}, \quad \Lambda \equiv \tilde\Lambda \, H_0^2,
\end{equation}
where $\hat\alpha$ and $\tilde\Lambda$ are dimensionless. Then, evaluating $q$ at $z=0$ (using the analytical expression for $H$ [Eq.~\eqref{Eq: H(z)}] and its derivative) yields the constraint:
\begin{equation}
    \delta > \frac{3 \Omega_{m_0} + 4 \Omega_{r_0}}{2 \hat\alpha}.
\end{equation}
For $\hat\alpha = 1$, thus ensuring the correct Bekenstein-Hawking limit when $\delta = 1$, and $\Omega_{m_0} = 0.3$ and $\Omega_{r_0} = 10^{-4}$ according to observations~\cite{Planck:2018vyg}, this reduces to:
\begin{equation}
    \delta \gtrsim 0.45.
\end{equation}
This condition ensures that the generalized entropy model predicts late-time cosmic acceleration as observed. However, as emphasized above, satisfying $q<0$ today does not guarantee a viable cosmology: the same corrections that drive acceleration also modify the cosmological dynamics toward the past, endangering the consistency of the early Universe.

\subsection{Cosmological Viability of Accelerated Solutions}

\begin{figure}
    \includegraphics[width=0.47\textwidth]{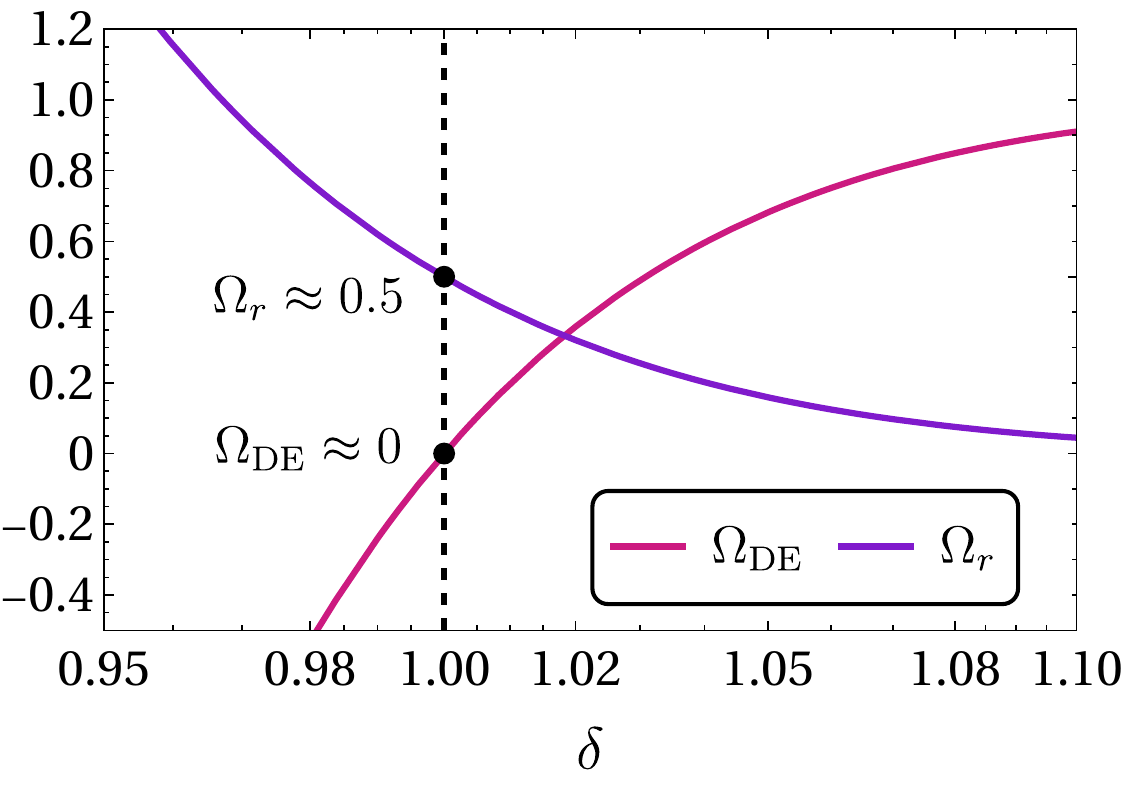}
    \caption{Density parameters $\Omega_{\DE}$ and $\Omega_r$ at $z \approx 3200$ as functions of the nonextensivity parameter $\delta$. The expected condition $\Omega_r \approx 0.5$ is satisfied only near $\delta \approx 1$, while values $\delta < 1$ lead to an overshoot of $\Omega_r$ and negative $\Omega_\DE$. For $\delta >1$, $\Omega_\DE$ prematurely dominates the energy budget eliminating the radiation-to-matter transition epoch.}
    \label{Fig: delta less 1}
\end{figure}

We now analyze the cosmological viability of the accelerated solutions identified in the previous section, which require $\delta \gtrsim 0.45$ to produce a negative deceleration parameter at present, i.e., at $z = 0$.

Equation~\eqref{Eq: DE Density Param} reveals that the Tsallis parameter must satisfy $\delta \neq 2$, since the expression diverges at this value. Moreover, for $\delta > 2$ the factor $(2 - \delta) < 0$, which drives the dark energy density parameter $\Omega_\DE$ to negative or complex values, rendering the model unphysical. At first glance, one might expect a physically allowed range $0.45 < \delta < 2$, but closer inspection shows that $\Omega_\DE$ also becomes negative throughout most of this interval. 

Figure~\ref{Fig: delta less 1} displays the density parameters $\Omega_\DE$ and $\Omega_r$ at a fixed redshift $z \approx 3200$, corresponding to the radiation–matter equality epoch, assuming $\Omega_{m_0} = 0.3$, $\Omega_{r_0} = 10^{-4}$, and $\hat{\alpha} = 1$. At this stage, standard cosmology predicts $\Omega_r \approx 0.5$ with negligible dark energy. However, this behavior is reproduced only within a narrow vicinity of the extensive limit $\delta = 1$. For $\delta < 1$, radiation rapidly overshoots unity, $\Omega_r \gtrsim 1$, and the excess is compensated by a negative $\Omega_\DE$, which is physically unacceptable. Conversely, for $\delta > 1$, the dark energy sector prematurely dominates the energy budget, erasing the expected sequence of radiation domination followed by matter domination and, finally, dark energy domination.

\begin{figure}
    \includegraphics[width=0.46\textwidth]{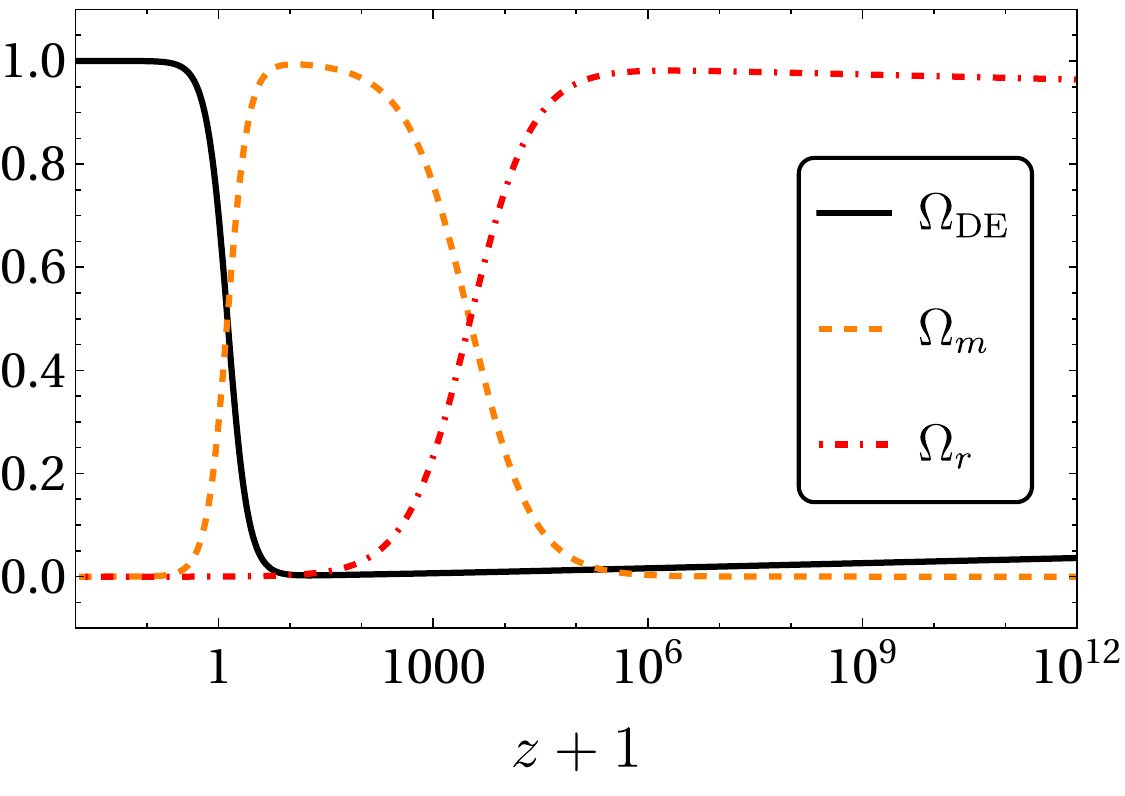}
    \caption{Cosmological evolution of $\Omega_{\DE}$, $\Omega_m$, and $\Omega_r$ for $\delta = 1.00037$, illustrating a standard sequence of radiation, matter, and dark energy domination. However, note that $\Omega_\DE$ tends to grow at early times.}
    \label{Fig: Densities}
\end{figure}

A more quantitative analysis shows that, for $\delta = 1.0021$ at $z \sim 3200$, we obtain:
\begin{align*}
\Omega_m \approx 0.4776,\quad \Omega_r \approx 0.4776,\quad \Omega_{\DE} \approx 0.0448,
\end{align*}
which is marginally consistent with the Big Bang Nucleosynthesis (BBN) constraint $\Omega_\DE(z \approx 3200) < 0.045$~\cite{Bean:2001wt}. However, for higher redshifts ($z > 3200$), the dark energy density quickly exceeds this upper limit. To guarantee compatibility with BBN up to $z \sim 10^{14}$, the highest redshift probed by standard Boltzmann solvers~\cite{Blas:2011rf}, we find the tighter bound:
\begin{equation}
\label{Eq: Allowed Range}
    1.00 \leq \delta < 1.00038.
\end{equation}

The CMB sets a less restrictive upper limit at $z \sim 50$, where $\Omega_\DE < 0.02$~\cite{Planck:2018vyg}. We find that this is saturated for $\delta = 1.0023$, with:
\begin{align*}
    \Omega_{m} \approx 0.9657,\quad \Omega_{r} \approx 0.0148,\quad \Omega_\DE \approx 0.0195,
\end{align*}
indicating a matter-dominated era at that redshift. However, this value of $\delta$ already violates the BBN bound, reinforcing that the allowed parameter space is tightly constrained around the extensive limit as in Eq.~\eqref{Eq: Allowed Range}.

The cosmological evolution within the ``viable range'' in Eq.~\eqref{Eq: Allowed Range} is illustrated in Fig.~\ref{Fig: Densities} for $\delta = 1.00037$. The standard thermal history is recovered: an initial radiation-dominated era, followed by matter domination, and a transition to late-time acceleration driven by dark energy. Note, however, that $\Omega_\DE$ tends to grow at the early stages in the evolution, contributing appreciably to the cosmic budget.  

\begin{figure}[t]
    \includegraphics[width=0.5\textwidth]{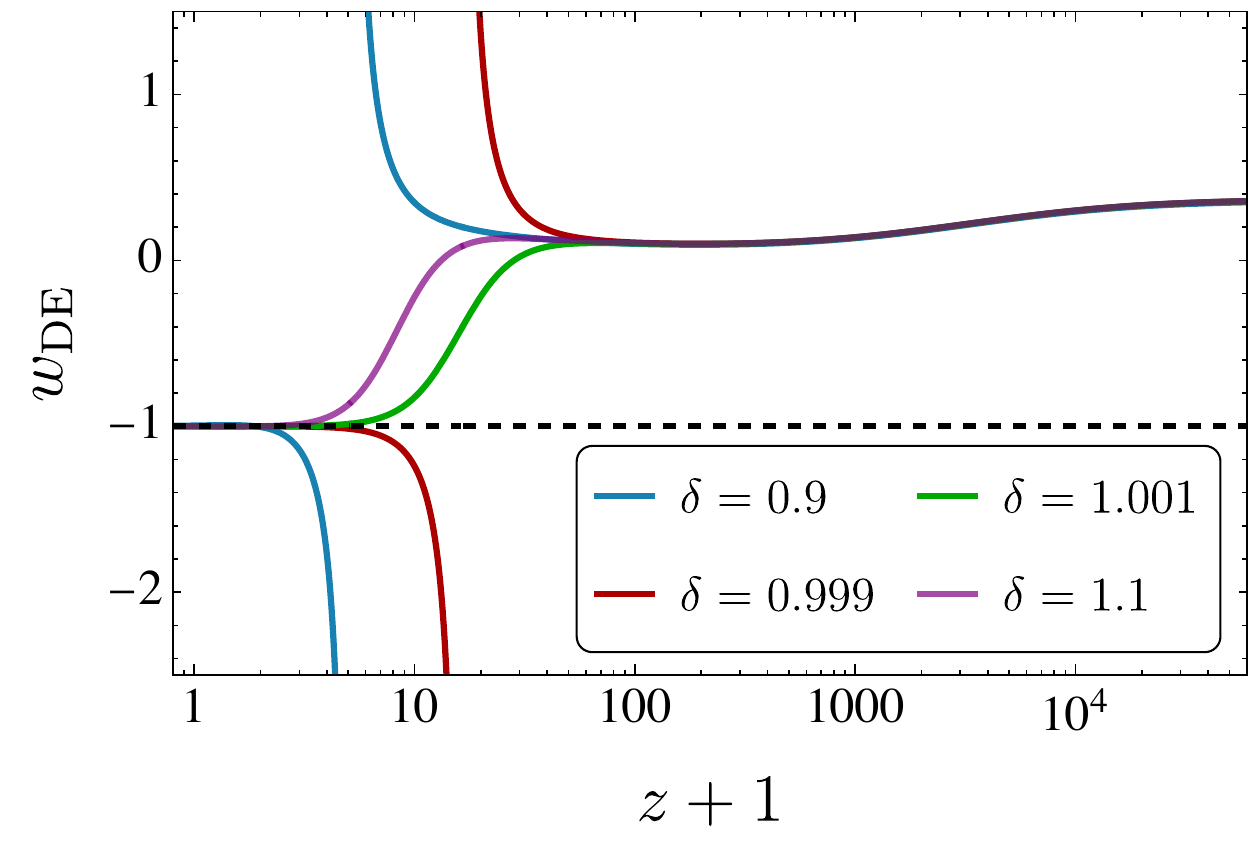}
    \caption{Evolution of the dark energy equation of state $w_{\DE}$ [Eq.~\eqref{Eq: wDE}] for different values of the non-extensivity parameter $\delta$. The black dashed line corresponds to the extensivity limit $\delta = 1$. For $\delta < 1$, $w_{\DE}$ diverges at some redshift, while in all cases the solutions converge to $w_{\DE} \to -1$ at late times.}
    \label{Fig: wDE}
\end{figure}

Aside from the issues related to the dark energy density, the equation of state $w_{\DE}$ also develops divergences when $\delta$ lies below the allowed range of Eq.~\eqref{Eq: Allowed Range}. As illustrated in Fig.~\ref{Fig: wDE}, for values such as $\delta = 0.9$ and $\delta = 0.999$, i.e., slightly below the extensivity limit $\delta = 1$, the denominator in Eq.~\eqref{Eq: wDE} vanishes at some redshift, leading to a divergence in $w_{\DE}$. Nevertheless, at late times the Universe asymptotically approaches a de Sitter phase in all cases, so that the divergence is not apparent when restricting the analysis to very low redshifts, $z \ll 10$. This explains why Refs.~\cite{Lymperis:2018iuz, Nojiri:2019skr}, which investigated the background dynamics only up to $z \sim 3$ and compared with supernovae data ($z \lesssim 2$), did not report this issue. As also shown in Fig.~\ref{Fig: wDE}, such divergences do not occur for $\delta > 1$, although this regime is already excluded by CMB and BBN constraints.

In the particular case $\Lambda = 0$, accelerated solutions would formally require $\delta < 1/2$. Nevertheless, this branch inherits the same pathologies identified for $\delta < 1$, thereby excluding it as a viable cosmological scenario.

In summary, the Tsallis horizon entropy with $\Lambda \neq 0$ is only consistent with observations for: 
$$
1.00 \leq \delta < 1.00038,
$$
a range essentially indistinguishable from the extensive case $\delta=1$. It is important to note, however, that $\Omega_{\DE}$ grows significantly at early times, contributing non-negligibly to the cosmic energy budget. This behavior is expected to disrupt the radiation-dominated era at sufficiently high redshift, well before $z \sim 10^{14}$. As we show in the next section, the apparently ``allowed range'' is therefore illusory, and the pathologies of the model are unavoidable.

Before closing this section, we stress that the introduction of an effective dark energy density and pressure in Eqs.~\eqref{Eq: DE density} and \eqref{Eq: DE pressure} should be understood as a convenient parametrization within the Cai–Kim thermodynamic framework, rather than as an assumption about the existence of a new physical fluid. Alternatively, one may regard Eq.~\eqref{Eq: First Friedman Eq} as defining a modified relation between the Hubble rate and the matter content, without introducing an effective energy-momentum component. As we shall show in the next section, the early-universe dynamics implied by the Tsallis entropy—and, in particular, the resulting constraints on the nonextensivity parameter  $\delta$—are independent of this interpretation, as they follow solely from the scaling of the modified Friedmann equation during the radiation-dominated era.

\section{Tsallis Cosmology as a Perturbative Extension of \texorpdfstring{$\Lambda$}{Lambda}CDM}
\label{Sec: Tsallis as a Perturbation}

In this section, we analyze the universe dynamics in the Tsallis cosmology, focusing on the radiation-dominated era. Tsallis cosmology is understood here as a perturbative deformation of the standard $\Lambda$CDM expansion history. The following discussion does not rely on interpreting the Tsallis-induced contribution as a dark energy component, but solely on the scaling of the Hubble rate with the radiation energy density implied by the modified Friedman equation. This relation can be obtained by substituting the effective dark energy density~$\rho_{\mathrm{DE}}$ from Eq.~\eqref{Eq: DE density} into the first Friedmann equation, Eq.~\eqref{Eq: First Friedman Eq}, and collecting all terms depending on the Hubble rate~$H$. This procedure yields a modified expansion law of the form:
\begin{equation}
\label{Eq: H_Tsallis}
    \frac{\alpha \, \delta}{2-\delta}\,H^{2(2-\delta)} = \frac{8\pi G}{3}\rho + \frac{\Lambda}{3}.
\end{equation}
The right-hand side corresponds to the standard Hubble parameter of $\Lambda$CDM, denoted as $H_{\Lambda}^{2}$. The cosmological constant appearing here is not necessarily the same as that in the $\Lambda$CDM model, but this distinction is irrelevant in the high-redshift regime considered in this section, where the contribution of $\Lambda$ is negligible. Thus one may rewrite Eq.~\eqref{Eq: H_Tsallis} as:
\begin{equation}
    H^2 = \left(\frac{2-\delta}{\alpha \, \delta} H_\Lambda^2\right)^{\frac{1}{2-\delta}}.
\end{equation}

Since the Tsallis model reduces to $\Lambda$CDM for $\delta=1$, it is natural to perform a series expansion around this value. Assuming $\hat{\alpha}=1$, the expansion reads:
\begin{equation}
H^{2} \approx H_{\Lambda}^2 + \delta H_{\Lambda, 1}^{2} \, (\delta-1) + \delta H_{\Lambda, 2}^{2}\,(\delta-1)^{2},
\end{equation}
where the first two expansion coefficients are:
\begin{align}
    \delta H_{\Lambda, 1}^{2} &\equiv
    H_{\Lambda}^{2}\left[\ln \left(\frac{H_{\Lambda}^{2}}{H_0^2}\right) - 2 \right], \\
    \delta H_{\Lambda, 2}^{2} &\equiv
    \frac{1}{2}H_{\Lambda}^{2}\left\{
    \left[\ln \left(\frac{H_{\Lambda}^{2}}{H_0^2}\right)\right]^{2}
    - 2 \ln \left(\frac{H_{\Lambda}^{2}}{H_0^2}\right) \right\}.
\end{align}
We truncate this expansion at second order in $(\delta-1)$. In this form, $H$ can be interpreted as the standard $\Lambda$CDM Hubble parameter plus perturbative corrections driven by the nonextensivity parameter $\delta$. Crucially, these corrections scale with $\ln(H_{\Lambda}^2/H_0^2)$ and thus grow large whenever $H_\Lambda \gg H_0$, i.e. at early times.

The impact on the radiation density parameter can be seen by writing:
\begin{equation}
\label{Eq: Omega_r}
    \Omega_r \propto \frac{\rho_r}{H_{\Lambda}^{2}} \left[1 + \frac{\delta H_{\Lambda, 1}^{2}}{H_\Lambda^2}(\delta-1) + \frac{\delta H_{\Lambda, 2}^{2}}{H_\Lambda^2}(\delta-1)^2 \right]^{-1},
\end{equation}
where we have used that $H_\Lambda^2$ dominates the background during the radiation era. In the standard limit $\delta=1$, one recovers $\Omega_r \to 1$ as $z \to \infty$, consistent with radiation domination. However, for $\delta \neq 1$ the bracket in Eq.~\eqref{Eq: Omega_r} deviates from unity by a factor that grows as $\ln(H_{\Lambda}^2/H_0^2) \propto \ln(1+z)^4$. Hence, even tiny departures from $\delta=1$ translate into large distortions of the radiation density at high redshift.

\begin{figure}
    \includegraphics[width=0.47\textwidth]{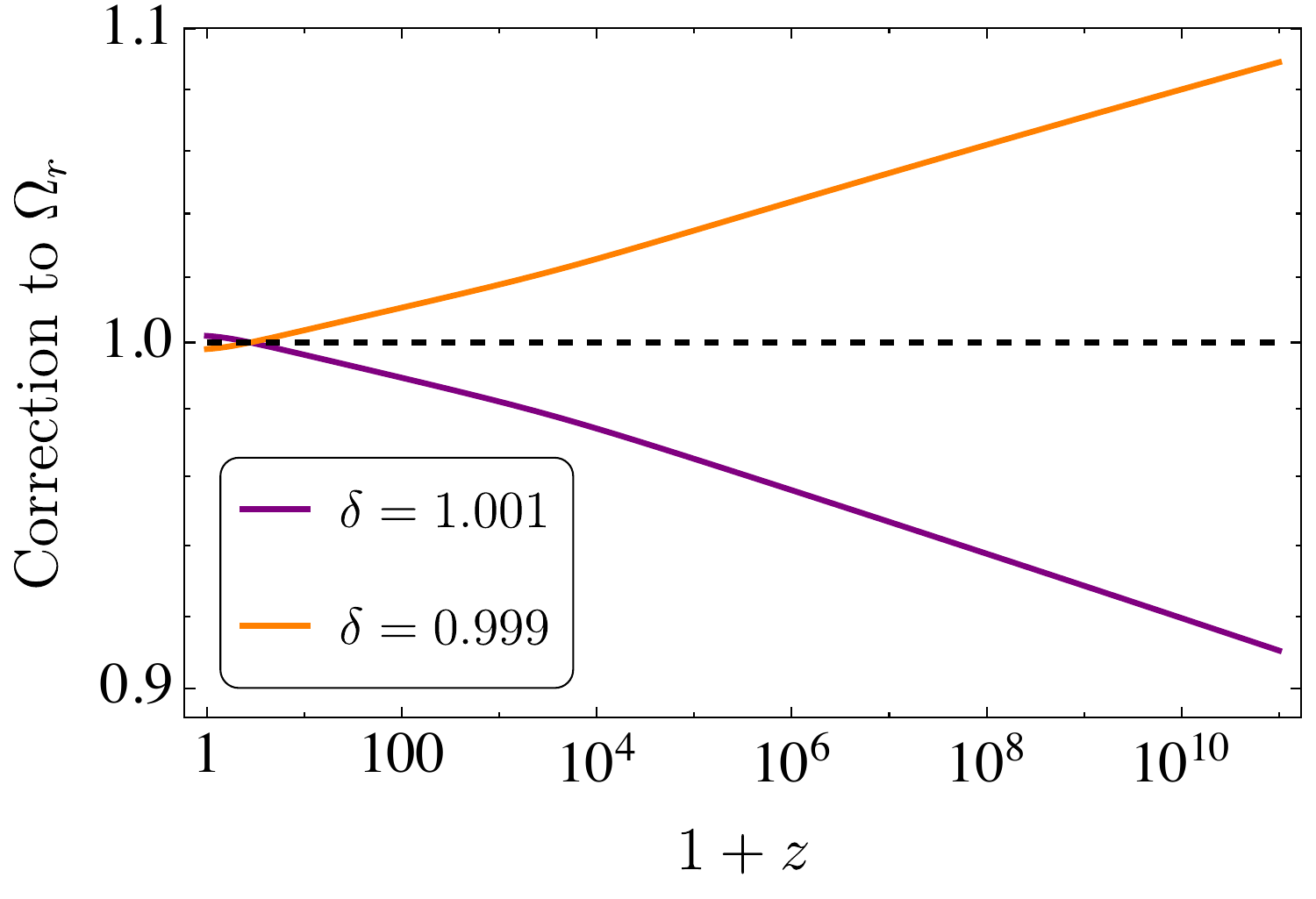}
    \caption{Deviation of the radiation density parameter from its standard $\Lambda$CDM evolution, as quantified by Eq.~\eqref{Eq: Omega_r}. For $\delta<1$ the correction enhances $\Omega_r$, forcing $\Omega_\DE<0$ at early times. For $\delta>1$ the correction suppresses $\Omega_r$, leading to an excessive early dark energy component. Both effects are direct consequences of the $H^{2(2-\delta)}$ scaling of Tsallis corrections.}
    \label{Fig: Correction}
\end{figure}

Figure~\ref{Fig: Correction} illustrates this behavior. For $\delta<1$, the correction factor becomes positive, yielding $\Omega_r>1$ and enforcing $\Omega_\DE<0$ through the Friedman constraint---an unphysical situation. For $\delta>1$, the correction suppresses $\Omega_r$ and instead requires a compensating dark energy component. This early dark energy fraction quickly exceeds the stringent bounds imposed by BBN and the CMB. In both cases, the underlying mechanism is the same: perturbative corrections that grow without bound as $H$ increases.

This perturbative analysis explains why cosmologically consistent solutions require $\delta$ to be indistinguishable from unity. Tsallis entropy introduces corrections that are negligible at late times, when $H$ is small, but diverge toward the past, destabilizing the standard radiation era. As a result, the only observationally viable limit of Tsallis cosmology is $\delta=1$, where the theory collapses back to $\Lambda$CDM. This provides a transparent physical interpretation of the issues found in Sec.~\ref{Sec: Cosmo Evolution}, and clarifies why the apparent freedom in $\delta$ is in fact illusory.

\section{Tsallis Holographic Dark Energy}
\label{Sec: THDE}

\subsection{Hubble Horizon Cut-off}

\begin{figure*}[t!]
\centering    
{\includegraphics[width=0.45\textwidth]{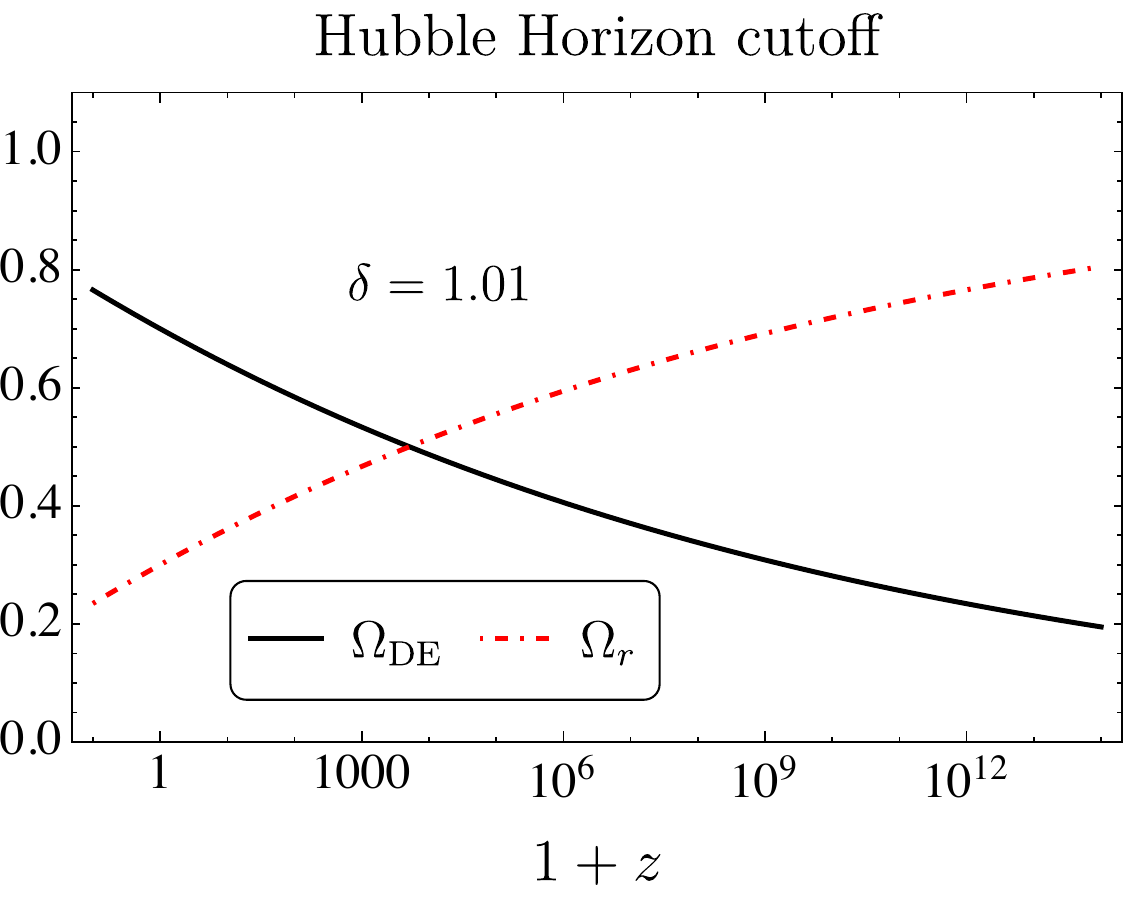}} \hfill
{\includegraphics[width=0.48\textwidth]{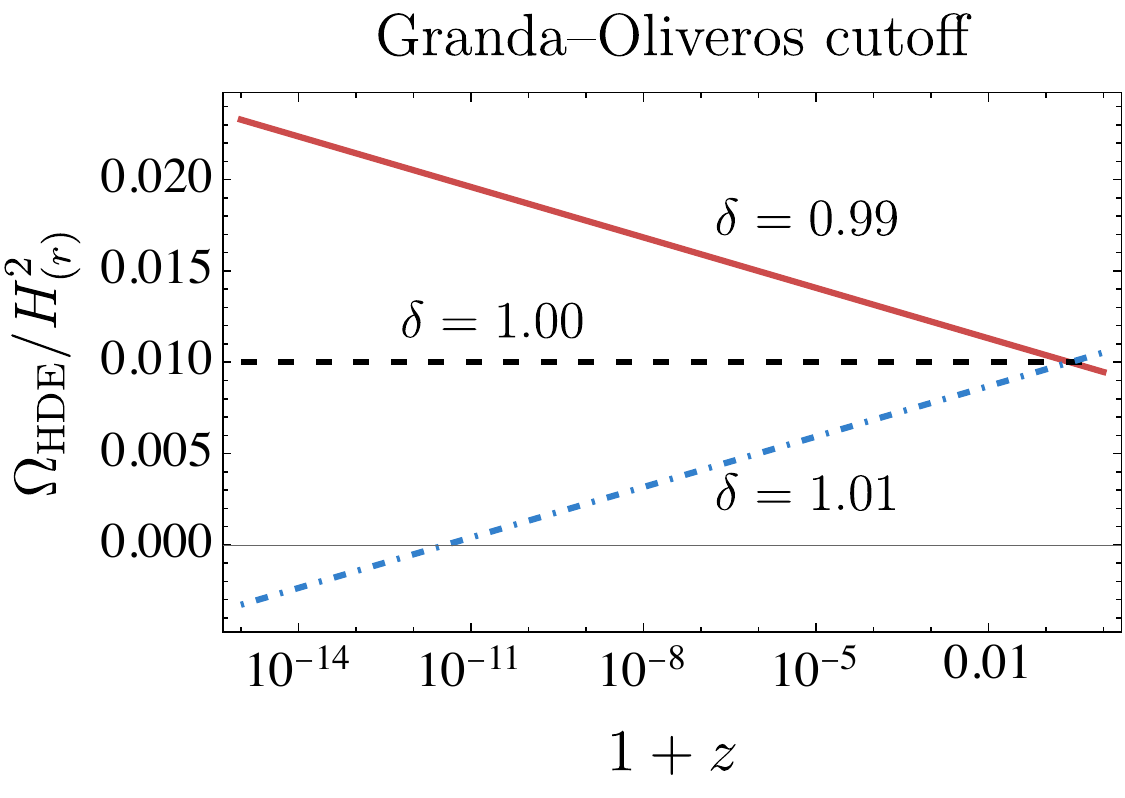}}
\caption{
\textbf{Left:} Evolution of $\Omega_r$ (red dot–dashed line) and the Tsallis holographic dark energy (HDE) density parameter $\Omega_\HDE$ (black solid line) for the Hubble horizon cutoff. For $\delta = 1.01$, even at extremely high redshift ($z = 10^{14}$), dark energy contributes about $20\%$ of the total cosmic budget, severely disrupting the radiation-dominated era. \textbf{Right:} Evolution of $\Omega_\HDE/\Omega_r$ for the Granda--Oliveros cutoff, considering $\alpha = 2\beta + 0.01$ and $\beta = 0.48$. When assuming radiation domination, this term should be near to zero. However, even for $\delta \approx 1$, this ``correction'' can be large enough to disrupt a proper radiation domination.}
\label{Fig: THDE}
\end{figure*}

An alternative implementation of Tsallis entropy in cosmology relies on the holographic principle, according to which the number of degrees of freedom scales with the boundary area rather than the volume of a system, subject to an infrared (IR) cutoff~\cite{Wang:2016och}. In the holographic dark energy (HDE) paradigm, the entropy $S$, the IR cutoff $L_\text{IR}$, and the ultraviolet (UV) cutoff $\Lambda$ are related by~\cite{Cohen:1998zx}:
\begin{equation}
    L_\text{IR}^3 \, \Lambda^3 \leq S^{3/4}.
\end{equation}
There are two common options for $L_\text{IR}$: the Hubble horizon and the Granda--Oliveros cutoff~\cite{Granda:2008dk, Granda:2008tm}. 

We firstly consider the Hubble horizon cutoff:
\begin{equation}
    L_\text{IR}^{(\text{H})} = 1/H.
\end{equation} 
Adopting Tsallis entropy $S_\text{T} = \gamma A^\delta$, with $\gamma$ a constant, the corresponding UV cutoff satisfies:
\begin{equation}
    \Lambda^4 \leq \left[\gamma (4\pi)^\delta\right] L^{2\delta - 4}_\text{IR}.
\end{equation}
This leads to the HDE energy density:
\begin{equation}
\label{Eq: HDE density}
    \rho_\HDE = B L^{2\delta - 4} = B \, H^{2(2 - \delta)},
\end{equation}
where $B$ is a constant. The first Friedman equation then becomes:
\begin{equation}
\label{Eq: HDE Friedman Eq}
    H^2 = \frac{8\pi G}{3}(\rho + \rho_\HDE),
\end{equation}
together with the standard continuity equations for a barotropic fluid and the HDE component:
\begin{align}
    \dot{\rho} + 3H(\rho + p) &= 0, \\
    \dot{\rho}_\HDE + 3H\rho_\HDE(1 + w_\HDE) &= 0,
\end{align}
where $w_\HDE$ is the HDE equation-of-state parameter. This construction was proposed in Refs.~\cite{Tavayef:2018xwx, Saridakis:2018unr} as a mechanism for late-time cosmic acceleration after a prior matter-dominated epoch. However, as we now show, it suffers from pathologies closely analogous to those of the Cai–Kim formulation when extending the cosmological evolution to the radiation-dominated epoch.

At high redshift, when nonrelativistic matter is negligible, Eq.~\eqref{Eq: HDE Friedman Eq} reduces to:
\begin{equation}
    H^2 = \frac{8\pi G}{3}(\rho_r + \rho_\HDE).
\end{equation}
Following Ref.~\cite{Tavayef:2018xwx}, the evolution of the HDE density parameter $\Omega_\HDE$ obeys:
\begin{equation}
    \frac{\dd \Omega_\HDE}{\dd \ln a} = 4(\delta - 1)\,\Omega_\HDE \left[\frac{1 - \Omega_\HDE}{1 - (2 - \delta)\Omega_\HDE}\right],
\end{equation}
with solution
\begin{equation}
    \Omega_\HDE \,[1 - \Omega_\HDE]^{1 -\delta} = C\, a^{4(\delta - 1)},
\end{equation}
where $C$ is determined by the initial conditions.

The left panel of Fig.~\ref{Fig: THDE} shows the evolution of $\Omega_r$ and $\Omega_\HDE$ for $\delta = 1.01$. Even at $z = 10^{14}$, the HDE component contributes nearly $20\%$ of the total energy density, clearly contradicting the standard radiation-dominated history. Therefore, a complete radiation era cannot be achieved within the allowed bounds derived in Ref.~\cite{Alvarez:2019ues}.

Interestingly, the pathology worsens as $\delta \to 1$: while in the Cai–Kim case deviations from $\delta = 1$ destabilize the early Universe, in the HDE scenario the limit $\delta \simeq 1$ itself is most problematic, as the HDE contribution remains non-negligible deep into the radiation era. This qualitative difference arises from the structure of the Friedman equation: in Cai–Kim cosmology, Tsallis corrections modify the geometric side [cf.~Eq.~\eqref{Eq: H_Tsallis}], whereas in the HDE framework they enter as an explicit energy density [cf.~Eqs.~\eqref{Eq: HDE density}--\eqref{Eq: HDE Friedman Eq}].

\subsection{Granda--Oliveros Cut-off}

A model employing the future event horizon as length scale can reproduce the observed late-time acceleration, but it suffers from the so-called causality problem—dark energy at present would depend on the future evolution of the scale factor~\cite{Li:2004rb}. The Granda--Oliveros (GO) cutoff was proposed to overcome this issue on purely dimensional grounds. It is defined as~\cite{Granda:2008dk, Granda:2008tm}:
\begin{equation}
    L_\text{IR}^{(\text{GO})} \equiv (\tilde\alpha H^2 + \tilde\beta \dot{H})^{-1/2},
\end{equation}
where $\tilde\alpha$ and $\tilde\beta$ are dimensionless constants. The first Friedman equation then reads:
\begin{align}
   3 m_\text{P}^2 H^2 &= \rho_r + \rho_m + \rho_\text{HDE}, \\
   \rho_\text{HDE} &\equiv 3 m_\text{P}^2\left( \alpha H^2 + \beta \dot{H} \right)^{2-\delta},
\end{align}
with $\alpha$ and $\beta$ dimensionful constants. 

At high redshift, neglecting matter, the Friedman equation reduces to $1 = \Omega_r + \Omega_\text{HDE}$, with:
\begin{equation}
    \Omega_\text{HDE} \equiv H^{-2} \left( \alpha H^2 + \beta \dot{H} \right)^{2-\delta}.
\end{equation}
A consistent radiation era requires $H^2_{(r)} \approx H_0^2 \Omega_{r0} a^{-4}$ and thus $\Omega_\text{HDE} \approx 0$. Expanding $\Omega_\text{HDE}/H^2_{(r)}$ around the extensivity limit $\delta = 1$, we find:
\begin{align}
    \frac{\Omega_\text{HDE}}{H^2_{(r)}} &= \left( \alpha - 2\beta \right) \nonumber \\
    &- \left( \alpha - 2\beta \right)(\delta -1) \ln \left[ \frac{H_0^2 \Omega_{r0}}{a^4} \left( \alpha - 2\beta \right) \right] \nonumber \\
     &+ \mathcal{O}\left[(\delta-1)^2\right].
\end{align}
This correction vanishes only for $\alpha = 2\beta$, as also noted in Ref.~\cite{Cardona:2022pwm} for the HDE model considering the Bekenstein--Hawking relation instead.

Since $\alpha < 2\beta$ would make the logarithm ill-defined, the only viable choice is $\alpha \gtrsim 2\beta$. Moreover, only for $\delta = 1$ does the correction remain constant in time. For $\delta < 1$, it grows as $\ln(1 + z)^4$, while for $\delta > 1$ it decreases with the same dependence, leading in both cases to an excessive early dark energy contribution. This behavior, shown in the right panel of Fig.~\ref{Fig: THDE}, prevents a consistent radiation-dominated epoch and closely mirrors the results obtained for Barrow holographic dark energy~\cite{Oliveros:2022biu}, which is mathematically related to Tsallis HDE though based on different conceptual approaches.\footnote{The Barrow entropy, given by $S_\text{B} \propto A^{1 + \Delta/2}$ with $\Delta$ the deformation parameter, can be seen as a analogous realization of Tsallis entropy when $\Delta = 2(\delta - 1)$. In particular, Barrow cosmology has been extensively studied in the literature~\cite{Luciano:2022ffn, Luciano:2023wtx, DiGennaro:2022grw, Leon:2021wyx}.}

Furthermore, Ref.~\cite{Oliveros:2022biu} showed that Barrow HDE can disrupt a fully matter-dominated era, with dark energy contributing up to $\sim 30\%$ of the total energy density at $z \simeq 10$ for small deformation parameters $\Delta$, corresponding to $\delta \approx 1$ in the Tsallis HDE formulation. By applying the same reasoning to a matter-dominated background ($H^2_{(m)} = H_0^2 \Omega_{m0} a^{-3}$), we find:
\begin{align}
    \frac{\Omega_\text{HDE}}{H^2_{(m)}} &= \frac{1}{2}\left( 2\alpha - 3\beta \right) \nonumber \\
    &- \frac{1}{2}\left( 2\alpha - 3\beta \right)(\delta -1) \ln \left[ \frac{H_0^2 \Omega_{m0}}{2a^3} \left( 2\alpha - 3\beta \right) \right] \nonumber \\
     &+ \mathcal{O}\left[(\delta-1)^2\right].
\end{align}
which vanishes only when $2\alpha = 3\beta$, in agreement with Ref.~\cite{Cardona:2022pwm}.
In general, assuming $\alpha \gtrsim 2\beta$ yields a negligible dark energy fraction during the radiation era but a significant one during matter domination. Conversely, if $2\alpha \gtrsim 3\beta$, dark energy remains subdominant throughout the matter-dominated epoch, while becoming non-negligible during radiation domination.

Interestingly, similar late-time scaling behavior—where dark energy evolves as pressureless matter—has been observed in interacting dark sector models~\cite{Copeland:1997et, Amendola:1999er, Copeland:2004qe, Orjuela-Quintana:2021zoe, Orjuela-Quintana:2022jrg, Patil:2023rqy}. Such ``clustering dark energy'' scenarios have been proposed to address the coincidence problem and to ease the $\sigma_8$ tension~\cite{Gonzalez:2018rop, Yang:2022csz, Jimenez:2024lmm, Benetti:2024dob}. However, within Tsallis HDE with the Granda–Oliveros cutoff, this scaling is not purely dynamical and it requires a delicate fine-tuning of the parameters $\alpha$ and $\beta$ to reproduce a cosmological evolution consistent with the expected high-redshift dynamics, and thus adding an extra ``coincidence'' problem.

\section{Conclusions}
\label{Sec: Conclusions}

In this work, we have performed a systematic assessment of Tsallis cosmology, analyzing both the Cai–Kim thermodynamic approach and the Tsallis holographic dark energy (HDE) scenario. In both formulations, the key ingredient is the nonextensivity parameter $\delta$, which modifies the Friedman equations by introducing terms proportional to $H^{2(1-\delta)}$. Our analysis shows that these corrections inevitably destabilize the early Universe, regardless of how close $\delta$ is to the extensive limit.

In the case of the Cai-Kim approach, for $\delta < 1$, the effective dark energy sector becomes negative, forcing the radiation density parameter $\Omega_r$ above unity to satisfy the Friedman constraint. For $\delta > 1$, the situation is reversed: the early dark energy fraction grows excessively, spoiling the standard radiation-dominated epoch required for big bang nucleosynthesis (BBN) and the formation of the cosmic microwave background (CMB). Quantitatively, consistency with early-Universe bounds requires
\[
1.00 \leq \delta < 1.00038,
\]
in agreement with previous constraints in the literature~\cite{Shahhoseini:2025sgl, DAgostino:2019wko, Leon:2021wyx, Luciano:2022ffn}.

We trace this pathology to the fact that the Tsallis deformation does not admit a perturbative interpretation around the predictions of the standard cosmological model. The expansion history cannot be described as $\Lambda$CDM plus small, controlled corrections, because even infinitesimal deviations $\delta - 1 \ll 1$ lead to qualitative failures in early-time cosmology. The model therefore reduces, in practice, to exact $\Lambda$CDM, as any physical departure is ruled out by consistency with the radiation era.

A closely analogous instability arises in the holographic formulation. When the Hubble horizon is adopted as the infrared cutoff, the Tsallis entropy induces a dark energy density $\rho_\HDE \propto H^{2(2-\delta)}$. In this case, the problem becomes even more pronounced: as $\delta \to 1$, the HDE contribution remains non-negligible deep into the radiation era, accounting for $\mathcal{O}(10\%)$ of the total energy density even at $z \sim 10^{14}$. Thus, although the mathematical structure of the Friedman equations differs between the Cai–Kim and HDE cases, both formulations collapse under the same mechanism—corrections involving powers of $H$ that grow uncontrollably toward the past.

When the Granda--Oliveros cutoff is adopted, the model can in principle be stabilized by fine-tuning the parameters such that $\alpha \gtrsim 2\beta$, preventing the breakdown of radiation domination. However, this choice leads to a controversial large dark energy contribution during the matter-dominated epoch, which might appear phenomenologically interesting in some contexts but generally casts doubt on the overall consistency of the model.

In summary, Tsallis horizon entropy—whether implemented through Cai–Kim thermodynamics or holographic dark energy—raises concerns about its viability as an alternative to the standard cosmological model. Its incorporation into the Friedman equations seems to compromise the standard thermal history, suggesting that only the extensive limit remains fully consistent with $\Lambda$CDM.

More broadly, our results highlight the importance of dynamical consistency and cosmological viability tests when evaluating nonextensive entropy formalisms as possible explanations for the Universe’s dynamics. Future work should explore whether this pathology is specific to Tsallis entropy or reflects a more general feature of nonextensive horizon thermodynamics. In particular, it will be crucial to investigate other generalized entropy frameworks—such as Kaniadakis~\cite{Sadeghnezhad:2021ekw, Lymperis:2021qty, Hernandez-Almada:2021rjs}, Rényi~\cite{Naeem:2022jdq}, and related proposals~\cite{Nojiri:2021iko, Nojiri:2022nmu, Nojiri:2023wzz, Nojiri:2024zdu, Odintsov:2025sew}—under the same early- and late-time cosmological tests, to establish whether any of these nonextensive entropy scenarios can yield a dynamically appealing cosmology.

\begin{acknowledgments}
This work was supported by Vicerrectoría de Investigaciones - Universidad del Valle Grant No. 71396. \\
\end{acknowledgments}

\bibliographystyle{apsrev4-2}
\bibliography{bibliography}

\end{document}